\documentclass[12pt]{article}
\usepackage{amsmath}
\begin{document}
\def\thefootnote{\fnsymbol{footnote}}
\begin{flushright}
KANAZAWA-09-04  \\
April, 2009
\end{flushright}
\vspace{ .7cm}
\vspace*{2cm}
\begin{center}
{\LARGE\bf Reconciliation of CDM abundance and $\mu\rightarrow e\gamma$
in a radiative seesaw model}\\
\vspace{1 cm}
{\Large Daijiro Suematsu}\footnote{e-mail:~suematsu@hep.s.kanazawa-u.ac.jp},
{\Large Takashi Toma}\footnote{e-mail:~t-toma@hep.s.kanazawa-u.ac.jp}
{\Large and Tetsuro Yoshida}\footnote{e-mail:~yoshida@hep.s.kanazawa-u.ac.jp}
\vspace*{1cm}\\
{\itshape Institute for Theoretical Physics, Kanazawa University,\\
        Kanazawa 920-1192, Japan}\\
\end{center}
\vspace*{1cm}
{\Large\bf Abstract}\\
We reexamine relic abundance of a singlet fermion as a CDM candidate, 
which contributes to the neutrino mass generation through radiative 
seesaw mechanism. We search solutions for Yukawa couplings and
the mass spectrum of relevant fields to explain neutrino oscillation data. 
For such solutions, we show that an abundance of a lightest singlet fermion
can be consistent with WMAP data without conflicting with both bounds of 
$\mu\rightarrow e\gamma$ and $\tau\rightarrow \mu\gamma$. 
This reconciliation does not need any modification of the original 
radiative seesaw model other than by specifying flavor structure of
Yukawa couplings and taking account of coannihilation effects.

\newpage
\section{Introduction}
Observation of nonzero neutrino masses \cite{oscil} and the
existence of dark matter \cite{wmap} gives big impact on the 
study of physics beyond the standard model (SM).
Elucidation of the origin of neutrino masses and dark matter is now 
one of the biggest subjects in this field.
Evidence for small neutrino masses suggests some additional structure 
in the SM as ones required in the seesaw mechanism \cite{seesaw}. 
On the other hand, in order to explain the abundance of dark matter, 
we need some symmetry to guarantee the stability of a dark matter candidate. 
The most famous example of such symmetry is $R$ parity in supersymmetric 
models. Even in non-supersymmetric models in general, 
if $Z_2$ symmetry is required by some natural reason,
it can play the same role as $R$ parity 
as long as the ordinary SM fields have its even charge.
From this point of view, there appears a very interesting idea that 
neutrino masses may be intimately related to the existence of cold dark matter.

If we consider that neutrino masses are generated radiatively under the
assumption that a model has $Z_2$ symmetry whose existence is justified
to forbid tree-level Dirac neutrino masses \cite{rad1}, this symmetry 
can guarantee the stability of a $Z_2$ odd neutral particle which may 
be cold dark matter (CDM).
In this direction the relation between neutrino masses and CDM has 
been studied in various articles \cite{scdm,cdmmeg,fcdm,ncdm}.
Since such scenarios require the introduction of new particles and 
interactions in the SM, however, it can induce dangerous 
effects in various phenomena as usual. In fact, it is suggested that
contradiction could appear between the strength of Yukawa couplings
to satisfy the required relic abundance of CDM and the bound for 
$\mu\rightarrow e\gamma$ as long as a singlet fermion is considered 
as a CDM candidate \cite{cdmmeg}. 
Some attempts have been proposed to overcome this fault by 
modifying the model \cite{ext}. 

In this paper we fix our target on a minimal model in \cite{rad1}, 
which can generate neutrino mass radiatively.
And we also confine our study into the case that a CDM candidate 
is one of the singlet fermions. 
We reanalyze whether its CDM abundance can be 
consistent with lepton flavor violating processes 
only by specifying detailed structure of the neutrino mass matrix 
and taking account of coannihilation effects.  
Our result will show that the CDM abundance can be consistent with 
lepton flavor violating processes even within 
the simplest radiative seesaw framework.

The paper is organized as follows. In the next section we define our model
briefly and discuss neutrino masses and mixing for the explanation of
neutrino oscillation data based on this model. 
In section 3 we show our result on both relic abundance of the CDM
candidate and constraints from the lepton flavor violating processes. 
We summarize the paper in section 4.  

\section{Neutrino mass due to radiative effects}
We consider a model which is an extension of the SM with 
an additional SU(2)$_L$ doublet scalar $\eta$ 
and three gauge singlet right-handed fermions $N_k$ \cite{rad1}. 
The model is also imposed by $Z_2$ symmetry to forbid tree-level
Dirac masses for neutrinos. We assign odd charge of this $Z_2$ 
symmetry to all of these new fields, although $Z_2$ even charge is assigned 
to all of the SM contents.
 
Lagrangian relevant to $N_k$ invariant under the SM gauge symmetry 
and this $Z_2$ symmetry are written as 
\begin{eqnarray}
{\cal L}_N&=& \left(i\overline{N_k}\gamma^\mu\partial_\mu P_R N_k\right)
 +\frac{1}{2}\left(M_k\overline{N^c_k}P_R N_k + M_k^\ast\overline{N_k}
P_L N_k^c\right)
-(h_{\alpha k}\overline{\ell_\alpha}\eta P_R N_k +{\rm h.c.}),
\label{int}
\end{eqnarray}
where $\ell_\alpha$ stands for a lepton doublet and a charged 
lepton mass matrix is assumed to be diagonalized. 
We note that $N_k$ can have mass terms invariant under the imposed symmetry.
For simplicity, these masses $M_k$ and Yukawa couplings $h_{\alpha k}$ 
are assumed to be real in the following discussion.
 
Scalar doublets $\Phi$ and $\eta$ have invariant scalar potential
\begin{eqnarray}
V&=&m_\Phi^2\Phi^\dagger\Phi+m_\eta^2\eta^\dagger\eta
+\frac{1}{2}\lambda_1(\Phi^\dagger\Phi)^2
+\frac{1}{2}\lambda_2(\eta^\dagger\eta)^2
+\lambda_3(\Phi^\dagger\Phi)(\eta^\dagger\eta)
+\lambda_4(\Phi^\dagger\eta)(\eta^\dagger\Phi) \nonumber \\
&+&\frac{1}{2}\lambda_5\left[(\Phi^\dagger\eta)^2
+{\rm h.c.}\right],
\end{eqnarray}
where $\Phi$ is the ordinary SM Higgs doublet.
If we assume that only $\Phi$ obtains a vacuum expectation value (VEV) 
such as $\langle\Phi^0\rangle=v$ but $\eta$ obtains no VEV, 
neutrinos cannot have tree-level Dirac masses.
However, neutrino masses can be generated radiatively through a 
one-loop diagram which has $\eta^0$ and $N_k$ in internal lines.
This radiative masses can be small as long as $\lambda_5$ is
sufficiently small.\footnote{Since we can introduce a new U(1) 
symmetry in case of $\lambda_5=0$, the smallness of $\lambda_5$ may be
considered as a natural assumption.}
Since $\lambda_5$ is assumed to be very small, masses of real and
imaginary parts of $\eta^0$ and also $\eta^\pm$ are considered to be
degenerate and they can be written as 
$m_0^2=m_\eta^2+(\lambda_3+\lambda_4)v^2$.
Thus, in the following discussion we use this $m_0^2$ as the mass of $\eta$.

Radiatively generated neutrino masses are expressed by using 
the Yukawa couplings $h_{\alpha k}$ and three mass scales $\Lambda_k$ as
\begin{equation}
({\cal M}_\nu)_{\alpha\beta}=\sum_{k=1}^3h_{\alpha k}h_{\beta k}\Lambda_k,
\label{mass1}
\end{equation}
where $\Lambda_k$ is defined by
\begin{equation}
\Lambda_k=\frac{\lambda_5 v^2}{8\pi^2M_k}I\left(\frac{M_k}{m_0}\right), 
\qquad
I(x)=\frac{x^2}{1-x^2}\left(1+\frac{x^2}{1-x^2}\ln x^2\right).
\label{mass2}
\end{equation}
By using this mass matrix, we now consider how to explain 
neutrino oscillation data.
Since it is known that neutrino oscillation data are well explained by
using the Maki-Nakagawa-Sakata (MNS) matrix\footnote{In this matrix 
$\sin\theta_{13}$ is
assumed to be zero, for simplicity. Since $\sin\theta_{13}$ is expected
to be very small, the present analysis is considered to be 
straightforwardly extended to the case with $\sin\theta_{13}\not=0$.} 
\begin{equation}
U=\left(\begin{array}{ccc}
\cos\theta & \sin\theta & 0 \\
-\frac{\sin\theta}{\sqrt 2} & \frac{\cos\theta}{\sqrt 2} & 
\frac{1}{\sqrt 2}\\
\frac{\sin\theta}{\sqrt 2} & -\frac{\cos\theta}{\sqrt 2} & 
\frac{1}{\sqrt 2}\\
\end{array}\right),
\label{mns}
\end{equation}
we assume that the mass matrix (\ref{mass1}) is diagonalized as 
$U^T{\cal M}_\nu U={\rm diag}(m_1,m_2,m_3)$.
Then, we find that the following diagonalization conditions 
should be satisfied:
\begin{eqnarray}
&&\sum_{k=1}^3\left(2h_{ek}^2\sin 2\theta +2\sqrt 2h_{ek}(h_{\mu k}-h_{\tau k})
\cos 2\theta-(h_{\tau k}-h_{\mu k})^2\sin 2\theta\right)=0, \nonumber \\
&&\sum_{k=1}^3h_{ek}\left(h_{\mu k}+h_{\tau k}\right)=0, \qquad 
\sum_{k=1}^3\left(h_{\mu k}-h_{\tau k}\right)\left(h_{\mu k}
+h_{\tau k}\right)=0.
\label{cond}
\end{eqnarray}
The mass eigenvalues are expressed as
\begin{eqnarray}
&&m_1=\sum_{k=1}^3\left(h_{ek}^2\cos^2\theta+\frac{1}{\sqrt 2}\sin 2\theta 
h_{ek}(h_{\tau k}-h_{\mu k})
+\frac{1}{2}\sin^2\theta\left(h_{\tau k}-h_{\mu k}\right)^2\right)\Lambda_k
, \nonumber \\
&&m_2=\sum_{k=1}^3\left(h_{ek}^2\sin^2\theta-\frac{1}{\sqrt 2}\sin 2\theta 
h_{ek}(h_{\tau k}-h_{\mu k})
+\frac{1}{2}\cos^2\theta\left(h_{\tau k}-h_{\mu k}\right)^2\right)\Lambda_k, 
\nonumber \\
&&m_3=\sum_{k=1}^3\frac{1}{2}\left(h_{\tau k}+h_{\mu k}\right)^2\Lambda_k. 
\label{mass}
\end{eqnarray} 
Here, among various solutions for the conditions (\ref{cond}),
we consider a simple solution such as
\begin{equation}
h_{ei}=0, \quad h_{\mu i}=h_{\tau i}; \quad h_{ej}\not=0,
\quad h_{\mu j}=-h_{\tau j},
\label{yukawa}
\end{equation}
where $i\not= j$ is assumed.\footnote{
If $h_{\mu k}=-h_{\tau k}$ is satisfied for all $k$, it is also a solution
for the diagonalization conditions. However, such a solution cannot satisfy 
neutrino oscillation data and then we do not consider this case here.}
This means either of $i$ or $j$ runs two values of $k=1,2,3$ such as
$i=1,2$ and $j=3$, for example.
By substituting this in the first condition in (\ref{cond}) we have
\begin{equation}
\tan\theta=-\frac{1}{\sqrt 2} \frac{h_{ej}}{h_{\tau j}}.
\label{theta}
\end{equation}
The mass eigenvalues (\ref{mass}) are rewritten as
\begin{eqnarray}
&&m_1=\left(h_{ej}\cos\theta+\sqrt 2h_{\tau j}\sin\theta\right)^2\Lambda_j
=0, \nonumber \\
&&m_2=\left(h_{ej}\sin\theta-\sqrt 2h_{\tau j}\cos\theta\right)^2\Lambda_j
=\frac{2h_{\tau j}^2}{\cos^2\theta}\Lambda_j, \nonumber \\
&&m_3=2h_{\tau i}^2\Lambda_i,
\end{eqnarray}
where the summation for $i$ and $j$ should be understood.
We use eq.~(\ref{theta}) in the last equality for
$m_{1,2}$.\footnote{This type of neutrino mass hierarchy induced from the
mass matrix (\ref{mass1}) has been considered to analyze neutrino
oscillation data in other context \cite{nmatrix}.}

Now we impose phenomenological requirements on the model. 
If we recall that the MNS matrix is given by $U$ defined in eq.~(\ref{mns}),
it is found that we can use $\sin^2\theta\simeq 0.33$ and 
$m_2^2\simeq 7.66\times 10^{-5}~{\rm eV}^2$ 
suggested by the solar neutrino and KamLAND data
and also $m_3^2\simeq 2.46\times 10^{-3}~{\rm eV}^2$ suggested by 
the atmospheric neutrino and K2K data \cite{oscil}. 
As this result, we obtain
\begin{equation}
h^2_{\tau j}\Lambda_j\simeq 2.9\times 10^{-3}~{\rm eV}, \qquad
h^2_{\tau i}\Lambda_i\simeq 2.5\times 10^{-2}~{\rm eV}, 
\label{osc}
\end{equation}
where the summation on $i$ and $j$ is abbreviated.  
Similar one-loop diagrams to the one for neutrino masses
contribute to the lepton flavor violating processes like 
$\ell_a\rightarrow \ell_b\gamma$. It gives the most severe constraint on
the model. Its branching ratio is estimated as \cite{bmeg}
\begin{equation}
Br(\ell_a\rightarrow \ell_b\gamma)=\frac{3\alpha}{64\pi(G_F m_0^2)^2}
\left|\sum_{k=1}^3h_{\ell_ak}h_{\ell_bk}
F_2\left(\frac{M_k}{m_0}\right)\right|^2,
\end{equation}
where $F_2(x)$ is given by
\begin{equation}
F_2(x)=\frac{1-6x^2+3x^4+2x^6-6x^4\ln x^2}{ 6(1-x^2)^4}.
\end{equation}
The present upper bounds for $Br(\mu\rightarrow e\gamma)$ 
and $Br(\tau\rightarrow \mu\gamma)$ 
are given as $1.2\times 10^{-11}$ \cite{expmeg} and $6.8\times 10^{-8}$
\cite{exptmg}, respectively.
If we use eq.~(\ref{yukawa}), these constraints can be written as 
\begin{eqnarray}
&&\left|h_{\tau j}^2F_2\left(\frac{M_j}{ m_0}\right)\right|
<9.8\times 10^{-4}\left(\frac{m_0}{500~{\rm GeV}}\right)^2, \nonumber\\
&&\left|h_{\tau i}^2F_2\left(\frac{M_i}{m_0}\right)
-h_{\tau j}^2F_2\left(\frac{M_j}{m_0}\right)\right|
<7.3\times 10^{-2}\left(\frac{m_0}{500~{\rm GeV}}\right)^2.
\label{meg}
\end{eqnarray}
If we discuss other phenomenological features of the model,
we should analyze them under the conditions (\ref{osc}) and (\ref{meg}).

To proceed with the study of CDM abundance in the next section,
it is convenient to classify possible cases for the 
relation between $M_k$ and $m_0$.
Since we assume $N_1$ to be a CDM candidate and we are interested in
the effect of coannihilation, $N_1$ is considered to be almost degenerate 
with other $Z_2$ odd fields. In such a situation, physically distinctive
cases may be classified as
$$  
{\rm (i)}~M_1~{^<_\sim}~ M_2 < M_3,~ m_0, \quad 
{\rm (ii)}~M_1~{^<_\sim}~ M_2,~ m_0 < M_3, \quad
{\rm (iii)}~M_1~{^<_\sim}~ m_0 < M_2,~ M_3.
$$
Although there is no logical correlation between Yukawa couplings 
and masses of singlet fermions without introducing some symmetry, 
here we only assume that the masses of the corresponding singlet 
fermions are equal if Yukawa couplings are equal in eq.~(\ref{yukawa}).  
We can identify important processes for determination of CDM abundance
under this assumption.
If we take $i=1,2$ and $j=3$, two possible cases (i) and (ii) should be
considered. 
In case (i) we need to take account of coannihilation between $N_1$ and $N_2$,
in which only Yukawa couplings are relevant to this process.
On the other hand, in case (ii) we take account of coannihilation 
among $N_1$, $N_2$ and $\eta$. Gauge interaction is expected to play 
an important role in this case.
If we take $i=1$ and $j=2,3$, 
the case (iii) with $M_2=M_3$ is a target for the investigation. 
In this case coannihilation between $N_1$ and
$\eta$ is expected to play a crucial role. 
Both gauge and Yukawa interactions are relevant to this case.
Although final states of coannihilation in the cases (ii) and (iii) 
can include antiproton, the case (i) can not include it but include only
lepton pairs. 
This aspect makes the case (i) interesting in the relation to 
the PAMELA $e^+$ and $\bar p$ data \cite{pamela}, and also the ATIC/PPB-BETS 
$(e^++e^-)$ data \cite{atic,ppb}. We will come back to this point later.
In the next section we will confine our study to this case.
Other cases will be discussed elsewhere.
 
\section{Coannihilation of the CDM candidate}
In this section we consider (co)annihilation of $N_1$ through Yukawa
couplings in the case (i).  
For the estimation of the relic abundance of $N_1$, 
we follow the method given in \cite{coann}, which is 
developed to take account of coannihilation effects.
If we introduce the dimensionless parameter $x$ as $x=M_1/T$, 
the decoupling temperature $T_f$ of $N_1$ can be estimated by using
effective cross section $\sigma_{\rm eff}$ and effective degrees of
freedom $g_{\rm eff}$ as
\begin{equation}
x_f=\ln\frac{0.038 g_{\rm eff} m_{\rm pl} M_1 \langle\sigma_{\rm eff}
|v_{\rm rel}|\rangle}{g_\ast^{1/2} x_f^{1/2}},
\end{equation}
where $v_{\rm rel}$ is the relative velocity of annihilating fields.
$\sigma_{\rm eff}$ and $g_{\rm eff}$ are defined as
\begin{eqnarray}
\sigma_{\rm eff}&=& 
\frac{g_{N_1}^2}{g_{\rm eff}^2}\sigma_{N_1N_1}+
2\frac{g_{N_1}g_{N_2}}{g_{\rm eff}^2}
\sigma_{N_1N_2}(1+\Delta)^{3/2}e^{-\Delta x}
+\frac{g_{N_2}^2}{g_{\rm eff}^2}\sigma_{N_2N_2}
(1+\Delta)^3e^{-2\Delta x}, \nonumber \\
g_{\rm eff}&=&g_{N_1}+g_{N_2}(1+\Delta)^{3/2}e^{-\Delta x},
\label{effcross}
\end{eqnarray}
where $m_{\rm pl}=1.22\times 10^{19}$~GeV and 
$\Delta$ is defined by $\Delta\equiv (M_2-M_1)/M_1$. 
If we define $a_{\rm eff}$ and $b_{\rm eff}$ by 
$\sigma_{\rm eff}|v_{\rm rel}|=a_{\rm eff}+b_{\rm eff}v_{\rm rel}^2$, 
thermally averaged cross section can be written as 
$\langle\sigma_{\rm eff}|v_{\rm rel}|\rangle=a_{\rm eff}+ 6 b_{\rm eff}/x$.
In the following
analysis, $\Delta\simeq 0$ is assumed since we consider the case (i).  
Thus, if we use this decoupling temperature $x_f$, 
the relic abundance can be estimated by 
\begin{equation}
\Omega h^2=\frac{1.07\times 10^9 x_f}{
g_\ast^{1/2}m_{\rm pl}({\rm GeV})(a_{\rm eff}+3b_{\rm eff}/x_f)}.
\label{relic}
\end{equation}  

(Co)annihilation proceeds via $t$-channel exchange of $\eta^0$ and $\eta^\pm$
through Yukawa interactions.
The final states are composed of only leptons $\mu$, $\tau$, $\nu_\mu$,
$\nu_\tau$ and their antiparticles. We note that antiproton is never produced.
The (co)annililation cross section of $N_{i_1}$ and $N_{i_2}$ is estimated as
\begin{eqnarray}
\sigma_{N_{i_1}N_{i_2}}|v_{\rm rel}|&=&\frac{1}{8\pi}
\frac{M_1^2}{(M_1^2+ m_0^2)^2}
\left[1 + \frac{m_0^4- 3m_0^2 M_1^2 -M_1^4}{3(M_1^2 
+m_0^2)^2}v_{\rm rel}^2 \right]
\sum_{\alpha,\beta}(h_{\alpha i_1}h_{\beta i_2}
-h_{\alpha i_2}h_{\beta i_1})^2 \nonumber \\
&+&\frac{1}{12 \pi}\frac{M_1^2(M_1^4+m_0^4)}{(M_1^2+m_0^2)^4}v_{\rm rel}^2
\sum_{\alpha,\beta}h_{\alpha i_1}h_{\alpha i_2}h_{\beta i_1}h_{\beta i_2},
\end{eqnarray}
where $i_1,i_2$ should be considered as 1 or 2.
As can be seen from this expression with $i_1=i_2=1$, 
the annihilation of $N_1$ occurs only through a $p$-wave channel.
On the other hand, coannihilation defined by $i_1\not= i_2$ can have 
$s$-wave contributions in general.
However, if we take account of the assumed conditions (\ref{yukawa}) 
in our model, we find that $s$-wave contributions cancel out and only
$p$-wave contributions remain.
Thus, the relevant cross section is found to be written as
\begin{equation}
\sigma_{N_{i_1}N_{i_2}}|v_{\rm rel}|=\frac{1}{3\pi}
\frac{M_1^2(M_1^4+m_0^4)}{(M_1^2+m_0^2)^4}
h_{\tau i_1}^2h_{\tau i_2}^2v^2_{\rm rel}. 
\label{cross}
\end{equation}
We have $a_{\rm eff}=0$ and
\begin{equation}
b_{\rm eff}=\frac{1}{12\pi m_0^2}\frac{r_1^2(1+r_1^4)}{(1+r_1^2)^4}
(h_{\tau 1}^2+h_{\tau 2}^2)^2,
\label{cs}
\end{equation}
where we define $r_1$ by $r_1=M_1/m_0$.
Applying this effective cross section to eq.~(\ref{relic}), we can
estimate the relic abundance of $N_1$. 

\input epsf
\begin{figure}[t]
\begin{center}
\epsfxsize=7cm
\leavevmode
\epsfbox{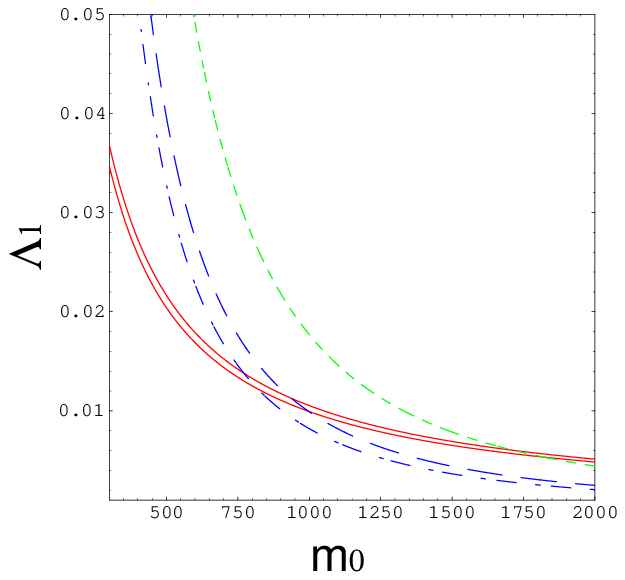}
\hspace*{5mm}
\epsfxsize=7cm
\leavevmode
\epsfbox{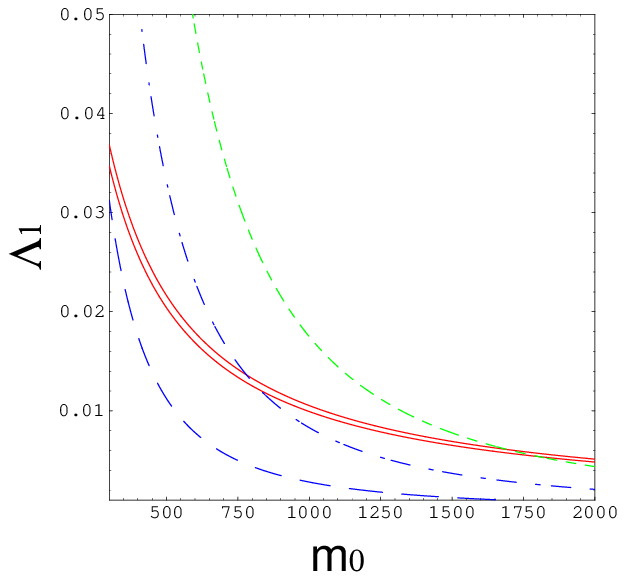}
\end{center}
\vspace*{-2mm}

{\footnotesize {\bf Fig.~1}~~Regions sandwiched by red solid lines   
satisfy the WMAP data $\Omega h^2=0.11\pm0.06$ for CDM abundance.
Blue dashed and blue dash-dotted lines show the bounds 
for $\mu\rightarrow e\gamma$ and $\tau\rightarrow \mu\gamma$, respectively. 
Green dotted lines represent contours for $\delta a_\mu=1.0\times 10^{-11}$. 
The values of $r_1$ and $r_3$ are fixed as 
$r_1=0.8$ in both graphs and $r_3=4$ and 10 in the left and right graphs. }
\end{figure}

\begin{figure}[t]
\begin{center}
\epsfxsize=7cm
\leavevmode
\epsfbox{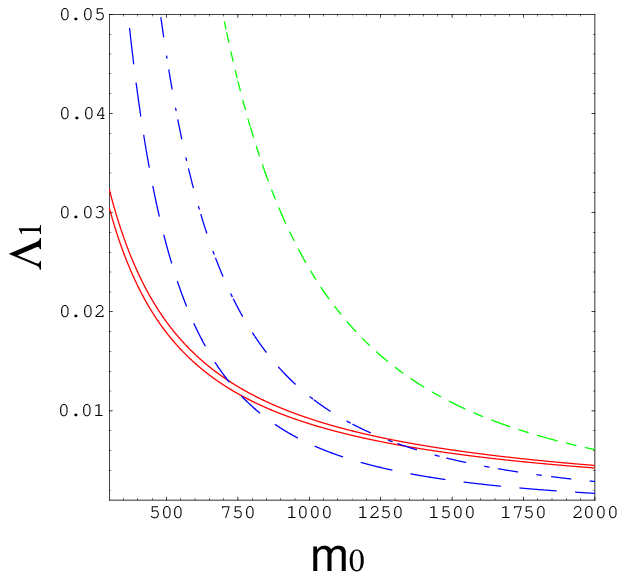}
\hspace*{5mm}
\epsfxsize=7cm
\leavevmode
\epsfbox{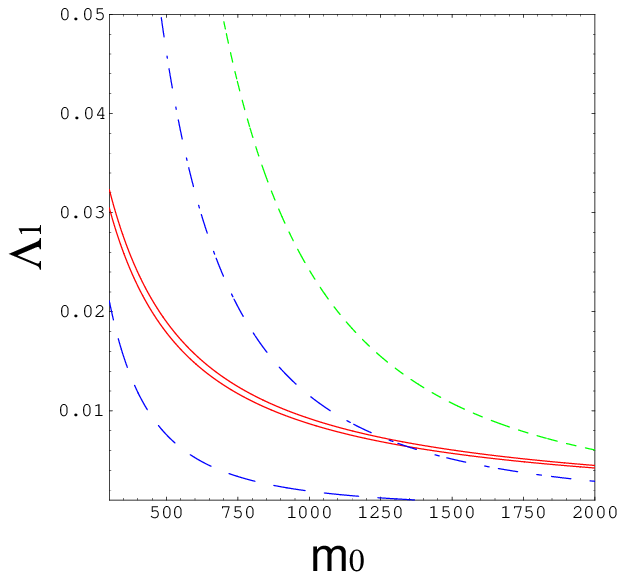}
\end{center}
\vspace*{-2mm}

{\footnotesize {\bf Fig.~2}~~The same figures as Fig.~1 for a different $r_1$. 
Although $r_1$ is fixed as $r_1=0.4$ in both graphs, $r_3$ is put as 4 
and 10 in the left and right graphs as in Fig.~1. }
\end{figure}

In the case (i), the constraints (\ref{osc}) obtained from 
the neutrino oscillation data are written as
\begin{equation}
h_{\tau 3}^2\Lambda_3 \simeq 2.9\times 10^{-3}~{\rm eV}, \qquad
(h_{\tau 1}^2+h_{\tau 2}^2)\Lambda_1\simeq 2.5\times 10^{-2}~{\rm eV}.
\label{osc1}
\end{equation}
If we use this relation in eq.~(\ref{cs}), the effective annihilation
cross section of $N_1$ can be expressed by using $\Lambda_1$ instead of
Yukawa couplings $h_{\tau 1}$ and $h_{\tau 2}$. 
As found from eq.~(\ref{meg}), the bound of $\mu\rightarrow e\gamma$ 
directly constrains only $h_{\tau 3}$ which is not relevant to the 
$N_1$ annihilation in this case. 
Since the relevant Yukawa couplings 
$h_{\tau 1}$ and $h_{\tau 2}$ are constrained by the bound of 
$\tau\rightarrow \mu\gamma$, this bound can be much more severe
constraint than that of $\mu\rightarrow e\gamma$. 
These constraints can also be written as conditions on $\Lambda_1$ as follows,
\begin{eqnarray}
&&\Lambda_1> 3.0 ~\frac{r_3I(r_1)}{r_1I(r_3)}F_2(r_3)
\left(\frac{500~{\rm GeV}}{m_0}\right)^2~{\rm eV}, \nonumber \\ 
&&\Lambda_1>\left(0.34F_2(r_1)-
4.0\times 10^{-2}\frac{r_3I(r_1)}{r_1I(r_3)}F_2(r_3)\right)
\left(\frac{500~{\rm GeV}}{m_0}\right)^2~{\rm eV}, 
\end{eqnarray}
where we define $r_3$ by $r_3=M_3/m_0$. We use a 
relation $\Lambda_3=(r_1I(r_3)/r_3I(r_1))\Lambda_1$ obtained from 
eq.~(\ref{mass2}) and also the conditions in eq.~(\ref{osc1}) 
in this derivation.

Using these results, we plot favorable regions in the 
$(m_0, \Lambda_1)$ plane by fixing the values of $r_1$ and $r_3$.
Since our considering case (i) corresponds to $r_1<1$ and $r_3\ge 1$, 
we fix these to some typical values.
We plot examples of the allowed regions for
$(r_1,r_3)=(0.8,4)$ and $(0.8,10)$ in Fig.~1 and
also for $(r_1,r_3)=(0.4,4)$ and $(0.4,10)$ in Fig.~2. 
In both figures, blue dashed and blue dash-dotted lines 
show the bounds for $\mu\rightarrow e\gamma$ and 
$\tau\rightarrow \mu\gamma$, respectively. 
The upper regions of both lines satisfy these constraints. 
Thin bands sandwiched by the red solid lines corresponds to region to 
realize the value of $\Omega h^2$ required by the WMAP data. 
Form this figure we find that the relic abundance of $N_1$ can be 
consistent with the WMAP data without conflicting the 
bounds of lepton flavor violating processes. 
Figs. 1 and 2 show that mass of $N_1$ should be larger than 700 GeV and
500 GeV for each case.
$\Lambda_3/\Lambda_1$ takes values of $O(1)$, for example, 
1.14 and 0.80 for $r_3=4$ and 10 in case of $r_1=0.8$, respectively. 
Although $\Lambda_1$ has rather small values $\sim 0.01$ 
in the allowed regions, Yukawa couplings $h_{\tau 1}$ and $h_{\tau 2}$ 
can be confirmed to be in the perturbative regions 
by taking account of eq.~(\ref{osc1}). 
Since these couplings contribute to $\tau \rightarrow \mu\gamma$,
this constraint can be much stronger than $\mu\rightarrow e\gamma$ 
as found from Figs.~1 and 2. However, we can find consistent solutions 
by fixing $r_1$ and $r_3$ suitably.

The reconciliation between the CDM abundance and the 
lepton flavor violating neutral processes can be shown to be 
accomplished even in the original radiative seesaw model without
substantial modification of the model.
In the present flavor structure of Yukawa couplings, 
the (co)annihilation of $N_1$ and $\mu \rightarrow e\gamma$ are induced by
the different ones, respectively. Yukawa couplings relevant to 
the (co)annihilation of $N_1$ contribute to $\tau \rightarrow \mu\gamma$
whose bound is much weaker than $\mu \rightarrow e\gamma$.
This feature makes their reconciliation possible by arranging the 
masses of singlet fermions so as to satisfy the requirements 
from the neutrino oscillation data.

Finally we give remarks on some predictions of the model.
Both direct and indirect detections of dark matter are crucial
to judge whether the considering model for dark matter is viable 
or not \cite{did,revdm}.  
Since the CDM candidate has couplings only to $\mu$ and $\tau$,
it can decay to these. Model independent analysis of the data of 
PAMELA and ATIC/PPB-BETS has been done in \cite{mindep,positron1}.
Ref.\cite{positron1} suggests 
that the best fit is obtained for $M\sim 1$~TeV with CDM
annihilating into $\mu^+\mu^-$ and a good fit is obtained for 
$M\sim 2$~TeV with CDM annihilating into $\tau^+\tau^-$.
It is interesting that this is consistent with our results obtained in
the present analysis. 
Encouraged by this result, we would like to add some qualitative 
arguments on the related subjects in our particular model. 

As shown in eq.~(\ref{cross}), $N_1$ annihilation cross section $\sigma|v|$ is 
dominated by $p$-wave contribution due to helicity suppression.
However, although $p$-wave contribution which has $v^2$ dependence dominates 
the annihilation cross section at freeze out time where $v\sim 0.2$, 
it is largely suppressed in the present Galaxy 
where $v\sim 10^{-3}$. This makes $s$-wave contribution relevant to PAMELA
anomaly rather than the $p$-wave contribution.
Since $s$-wave annihilation cross section can be estimated as 
$\sigma|v|\simeq \frac{h_{\alpha 1}^2}{8\pi}
\frac{m_f^2}{m_0^4(1+r_1^2)^2}$ where $m_f$ is the mass of final fermions,
$N_1$ annihilation in the Galaxy occurs mainly through 
$N_1N_1\rightarrow \tau^+\tau^-$.
If we use typical values of $m_0$, $h_{\tau 1}$ and $r_1$ obtained 
as the solutions consistent with the WMAP data in this paper, we find that 
the boost factor should be $O(10^6)$ or larger.\footnote{We note 
that $\sigma|v|\sim 10^{-23}~{\rm cm}^3/{\rm sec}$ is required to explain 
the positron excess in the PAMELA data.}
The model cannot induce this amount of enhancement for the
annihilation cross section in the present form.
For the explanation of this boost factor, there may be two
possibilities: (i) the model should be extended such that the relic 
$N_1$ has a large non-thermal component as discussed in \cite{twocomp},
for example,
or (ii) it should be explained by some astrophysical effects.
However, it seems difficult to obtain substantial effects by a simple
extension referred in case (i). 
Since we can not make the annihilation cross section of $N_1$ itself larger
preserving the features of the model, main effect should come from the 
increase of number density of the non-thermal component of relic $N_1$ 
as a result of the decay of other fields. However, it is severely 
constrained by the WMAP data and we have no freedom to obtain the 
large boost factor mentioned above.   

Bremsstrahlung from the charged fields associated to this annihilation
$N_1N_1\rightarrow \tau^+\tau^-$ yields diffuse photons. We may check the
model by comparing the flux of diffuse photon expected from this 
$N_1$ annihilation based on the PAMELA data with observations such as Hess 
and Fermi/GLAST. Such a model independent analysis is presented in
\cite{gamma}. 
Although it suggests that dark matter annihilation with $\tau^+\tau^-$ 
final states may be difficult to be consistent with diffuse photon data, 
the assumptions in that analysis seems not to be applied to our model. 
On the other hand, dark matter annihilation into final states composed of 
three fields such as $e^+e^-\gamma$ can be dominant processes if the 
annihilation cross section is helicity suppressed \cite{positron2}. 
The radio emission from synchrotron radiation and $\gamma$-ray emission
from inverse Compton scattering from the charged fields produced by the dark
matter annihilation may be also useful to discriminate the origin of
positron excess \cite{dphoton}.
Observational data of diffuse photon obtained in the Fermi/GLAST experiment 
may give us crucial hints for these \cite{fermi}.  
Anyway, detailed analysis of diffuse photon is necessary to check the
validity of the present model.   

Decay of $\tau^\pm$ also produces neutrino flux. If we use the PAMELA 
positron data, its flux can be roughly estimated as $O(10^{5\sim 6})$ 
GeV/(cm$^2\cdot$sec$\cdot$str) at the relevant neutrino energy. 
This flux is larger than neutrino flux expected from certain types of AGN
but smaller than the atmospheric neutrino flux \cite{nflux}.
This suggests that the $N_1$ annihilation is difficult to be detected 
through the observation of the neutrino flux on the Earth.   

Direct detection of $N_1$ is also an interesting subject. 
Although $\eta$ has the interaction shown in eq.~(\ref{int}) 
and no direct interaction with quarks, $N_1$ can be scattered 
by nuclei through one-loop effect with $Z$ boson exchange.
Since $N_1$ is a Majorana fermion, this effective interaction with quark
is expressed by an axial vector interaction 
$d_q\bar N_1\gamma_5\gamma_\mu N_1\bar q\gamma_5\gamma^\mu q$ with
$d_q\sim \frac{g_2^2h_{\tau 1}^2r_1^2}{(4\pi)^2m_W^2}T_{3q}$, which yields
spin dependent scattering.
If we use the parameters obtained in this paper, 
this spin dependent cross section is roughly estimated as 
$O(10^{-41})~{\rm cm}^2$.
This is much smaller than the present bound of spin dependent elastic 
scattering cross section for 
dark matter with $O(1)$~TeV mass \cite{direct}. 
Thus, it seems difficult to find this dark matter even in the next generation 
direct detection experiments.   
Studies related to these aspects of the similar model can also be found in 
\cite{positron3} although lepton number violating constraints are not
taken into account there.

In addition to these indirect and direct search of dark matter,
there may be some other phenomena which could show characteristic 
features of the model. 
The effective mass in the neutrinoless double $\beta$ decay is
given as a fixed value $m_{\rm eff}= 
\sqrt{\Delta m_{\rm sol}^2}\sin\theta_{\rm
sol}\simeq 2.9\times 10^{-3}$ eV, which is one order of magnitude below
the reach of near future experiments.
Present data for the magnetic dipole moment of muon shows 
discrepancy between a value predicted by the SM and the experimental 
result \cite{expg2}. In the present model there is
one-loop contribution to $\delta a_\mu$, which is estimated as \cite{bmeg}
\begin{eqnarray}
\delta a_\mu&=&\sum_{k=1}^3\frac{h_{\mu k}^2}{(4\pi)^2}
\frac{m_\mu^2}{m_0^2}F_2(r_k) \nonumber \\
&\simeq&\frac{7.1\times 10^{-12}}{\Lambda_1}\left(\frac{500~{GeV}}
{m_0}\right)^2
\left[ F_2(r_1)+0.12\frac{r_3I(r_1)}{r_1I(r_3)}F_2(r_3) \right].
\end{eqnarray}
A contour for $\delta a_\mu =1.0\times 10^{-11}$ is also plotted 
by a green dotted line in the figures. 
This shows that these values predicted by our model are two 
orders of magnitude smaller than  
$\delta a_\mu=(30.2\pm 8.7)\times 10^{-10}$ \cite{mg2}. 
In order to improve this situation for the $\delta a_\mu$,
additional contributions to $\delta a_\mu$ are required in our model. 
Such contributions may be obtained by embedding our scenario 
in supersymmetric models, in which an ordinary supersymmetric 
CDM candidate such as
the lightest neutralino does not dominate the required relic abundance. 
We will discuss such extensions elsewhere.

\section{Summary}
The radiative seesaw model considered in this paper 
is one of interesting possibilities to explain the origin of
neutrino masses. It can include a cold dark matter candidate as an
important ingredient of the neutrino mass generation.
However, the model has been considered to have a severe discrepancy 
between magnitude of Yukawa couplings required 
by the dark matter relic abundance and the suppression of lepton 
flavor violating neutral processes.
In this study we have proposed a new possibility to relax this tension 
within the original minimal radiative seesaw model without 
introducing additional interactions.
We have found that the model can overcome this problem simultaneously 
satisfying the conditions required by the neutrino oscillation data 
as long as the Yukawa couplings and also the mass hierarchy of the 
singlet fermions have appropriate structure. 
The present study shows that even the minimal radiative seesaw model 
can be an interesting candidate for models which relates neutrino masses 
to the existence of dark matter. 

The model may be relevant to the PAMELA
$e^+$ and $\bar p$ data, and also the ATIC/PPB-BETS $(e^++e^-)$ data. 
We have briefly presented qualitative observations on the detection of
the diffuse particles produced in the dark matter annihilation and the
direct search of this dark matter. Since these analyses are rough and
qualitative ones, we need more quantitative study to mention on the
predictions of the model in detail.
This scenario may play an important role in some supersymmetric 
models if it is embedded in the supersymmetric framework.
Although we have considered only the model with restricted 
coannihilation processes here, other cases are also expected to give 
interesting possibilities. These points may be worthy for further study
and will be discussed elsewhere.

\vspace*{5mm}
\noindent
This work is partially supported by a Grant-in-Aid for Scientific
Research (C) from Japan Society for Promotion of Science (No.17540246).


\newpage
\begin{thebibliography}{99}
\bibitem{oscil}SNO Collaboration, Q.~R~.Ahmad, {\it et al.},
	Phys. Rev. Lett. {\bf 89} (2002) 011301;
	Super-Kamiokande Collaboration, Y.~Fukuda, {\it et al.},
	Phys. Rev. Lett. {\bf 81} (1998) 1562; 
        KamLAND Collaboration, K.~Eguchi, {\it et al.}, 
        Phys. Rev. Lett. {\bf 90} (2003)
	021802; 
        K2K Collaboration, M.~H.~Ahn, {\it et al.},
	Phys. Rev. Lett. {\bf 90} (2003) 041801.    

\bibitem{wmap}WMAP Collaboration, D.~N.~Spergel {\it et al.},
	Astrophys. J. {\bf 148} (2003) 175.

\bibitem{seesaw} P.~Minkowski, Phys Lett. {\bf B67} (1977) 421;
	T.~Yanagida, in Proc. Workshop on Unified Theory and Baryon
	Number in the Universe, eds. O.~Sawada and A.~Sugamoto (KEK, Tsukuba,
	1979); M.~Gell-Mann, P.~Ramond and R.~Slansky, in Supergravity,
	eds. P. van Nieuwenhuizen and D.~Freedman (North-Holland, 
       Amsterdam, 1979) p.315.

\bibitem{rad1} E.~Ma, Phys. Rev. {\bf D73} (2006) 077301.

\bibitem{scdm}R.~Barbieri, L.~E.~Hall and V.~S.~Rychkov, Phys. Rev. {\bf
	D74} (2006) 015007; L.~Lepoz~Honorez, E.Nezri, J.~F.~Oliver and
	M.~H.~G.~Tytgat, JCAP {\bf 02} (2007) 28; M.~Gustafsson,
	E.~Lundstrom, L.~Bergstrom and J.~Edsjo, Phys. Rev. Lett. {\bf
	99} (2007) 041301.  

\bibitem{cdmmeg} J.~Kubo, E.~Ma and D.~Suematsu, Phys. Lett. 
{\bf B642} (2006) 18.

\bibitem{fcdm} L.~M.~Krauss, S.~Nasri and M.~Trodden, Phys. Rev. {\bf
	D67} (2003) 085002; D.~Aristizabal Sierra, J.~Kubo, D.~Restrepo,
	D.~Suematsu and O.~Zepata, Phys. Rev. {\bf D79} (2009) 013011;
	M.~Aoki, S.~Kanemura and O.~Seto, Phys. Rev. Lett. {\bf 102} 
(2009) 051805.

\bibitem{ncdm} M.~Lattanzi and V.~W.~F.~Valle, Phys. Rev. Lett. {\bf 99}
	(2007) 121301; C.~Boehm, Y.~Farzan, T.~Hambye, S.~Palomares-Ruiz
	and S.~Pascoli, Phys. Rev. {\bf D 77} (2008) 043516; E.~Ma,
	Phys. Lett. {\bf B662} (2008) 49. 

\bibitem{ext} J.~Kubo and D.~Suematsu, Phys. Lett. {\bf B643} (2006) 336;
K.~S.~Babu and E.~Ma, Int. J. Mod. Phys. {\bf A23} (2008) 1813; 
D.~Suematsu, Eur. Phys. J. {\bf C56 } (2008) 379; E.~Ma and D.~Suematsu, 
Mod. Phys. Lett. {\bf A24} (2009) 583. 

\bibitem{nmatrix} D. Suematsu, Phys. Lett. {\bf B392} (1997) 413;
	Prog. Theor. Phys. {\bf 99} (1998) 483; Int. J. Mod. Phys. {\bf
	A15} (2000) 3967; Prog. Theor. Phys. {\bf 106} (2001) 587.

\bibitem{bmeg} E.~Ma and M.~Raidal, Phys. Rev. Lett. {\bf 87} (2001) 011802.

\bibitem{expmeg}MEGA Collaboration, M.~L.~Brooks, {\it et al.}, 
Phys. Rev. Lett. {\bf 83} (1999) 1521.

\bibitem{exptmg}BABAR Collaboration, B.~Aubert, {\it et al.}, 
Phys. Rev. Lett. {\bf 95} (2005) 041802.

\bibitem{pamela} O.~Adriani {\it et al.}, arXiv:0810.4995 [astro-ph].

\bibitem{atic}J.~Chang {\it et al.}, Nature {\bf 456} (2008) 362.

\bibitem{ppb}S.~Torii {\it et al.}, arXiv:08090760 [astro-ph].

\bibitem{coann} K.~Griest and D.~Seckel, Phys. Rev. {\bf D43} (1991) 3191.

\bibitem{did}M.~Beltran, D.~Hooper, E.~W.~Kolb and Z.~A.~C.~Krusberg,
	arXiv:0808.3384 [hep-ph].

\bibitem{revdm}For a review, see for example, 
G.~Jungman, M.~Kamionkowski and K.~Griest,
	Phys. Rept. {\bf 267} (1996) 195; G.~Bertone, D.~Hooper and
	J.~Silk, Phys. Rept. {\bf 405} (2005) 279.  

\bibitem{mindep}V.~Bager, W.-Y.Keung, D.~Marfatia and G.~Shaughnessy,
	Phys. Lett. {\bf B672} (2009) 141; I.~Cholis, L.~Goodenough,
	D.~Hooper, M.~Simet and N.~Weiner, arXiv:0809.1683 [hep-ph].

\bibitem{positron1}M.~Cirelli, M.~Kadastik, M.~Raidal and A.~Strumia,
	arXiv:0809.2409. 

\bibitem{twocomp}M.~Fairbairn and J.~Zupan, arXiv:0810.4147 [hep-ph].

\bibitem{gamma}G.~Bertone, M~Cirelli, A.~Strumia and M.~Taoso, 
arXiv:0811.3744 [astro-ph].

\bibitem{positron2}L.~Bergstro\"m, T.~Bringmann and J.~Edsj\"o,
	arXiv:0808.3725 [astro-ph].

\bibitem{dphoton}J.~Zhang, X.-j. Bi, J.~Liu, S.-M. Liu, P.-f. Yiu,
	Q.~Yuan and S.-h. Zhu, arXiv:0812.05222 [astro-ph].

\bibitem{fermi}E.~A.~Baltz, {\it et al.}, JCAP {\bf 0807} (2008) 013.

\bibitem{nflux}J.~K.~Becker, P.~L.~Biermann and W.~Rhode, 
Astropart. Phys. {\bf 23} (2005) 355.
 
\bibitem{direct}XENON Collaboration, J.~Angle, {\it et al.},  
Phys. Rev. Lett. {\bf 100} (2008) 021303; 
CDMS Collaboration, Z.~Ahmed, {\it et al.}, Phys. Rev. Lett. {\bf 102}
	(2009) 011301.

\bibitem{positron3}Q.-H.~Cao, E.~Ma and G.~Shaughnessy,
	arXiv:0901.1334 [hep-ph].

\bibitem{expg2}Muon $g-2$ Collaboration, G.~W.~Bennett {\it et al.},
	Phys. Rev. {\bf D73} (2006) 072003.

\bibitem{mg2}K.~Hagiwara, A.~D.~Martin, D.~Nomura and T.~Teubner,
	Phys. Lett. {\bf B649} (2007) 173.

\end{thebibliography}
\end{document}